\documentclass[10pt,conference]{IEEEtran}
\IEEEoverridecommandlockouts

\usepackage{cite}
\usepackage{url}
\usepackage{bbm}
\usepackage{amsmath,amsfonts} 
\usepackage{float}
\usepackage{graphicx}
\usepackage{algorithm}
\usepackage{algpseudocode}
\usepackage{siunitx}
\usepackage[labelfont=bf]{caption}
\usepackage{subcaption}
\usepackage{tikz}
\usepackage{pgfplots}
\usepackage{epstopdf}
\usepackage{multirow}
\usepackage[font=scriptsize]{caption}
\usepackage{soul}
\usepackage{color, colortbl}
\usepackage{array}
\usepackage[T1]{fontenc}
\usepackage[utf8]{inputenc}
\usepackage{mathptmx}
\usepackage{listings}
\usepackage{adjustbox}
\usepackage{makecell}

\usepackage{enumitem}
\usepackage{hyperref}
\usepackage{pifont}

\def\BibTeX{{\rm B\kern-.05em{\sc i\kern-.025em b}\kern-.08em
    T\kern-.1667em\lower.7ex\hbox{E}\kern-.125emX}}
    
\makeatletter
\def\endthebibliography{%
  \def\@noitemerr{\@latex@warning{Empty `thebibliography' environment}}%
  \endlist
}
\makeatother

\begin{document}

\title{A Public Key Infrastructure for 5G Service-Based Architecture}


\author{\IEEEauthorblockN{Ayush Kumar and Vrizlynn L.L. Thing}\\
\IEEEauthorblockA{Cyber Security Strategic Technology Centre \\
Singapore Technologies Engineering \\
Email: ayush.kumar@u.nus.edu, vriz@ieee.org}
}
			


\maketitle

\begin{abstract}
The 3GPP 5G Service-based Architecture (SBA) security specifications leave several details on how to setup an appropriate Public Key Infrastructure (PKI) for 5G SBA, unspecified. In this work, we propose \textsf{5G-SBA-PKI}, a public key infrastructure for secure inter-NF communication in 5G SBA core networks, where NF refers to Network Functions. \textsf{5G-SBA-PKI} is designed to include multiple certificate authorities (with different scopes of operation and capabilities) at different PLMN levels for certification operations and key exchange between communicating NFs, where PLMN refers to a Public Land Mobile Network. We conduct a formal analysis of \textsf{5G-SBA-PKI} with respect to the desired security properties using TAMARIN prover. Finally, we evaluate \textsf{5G-SBA-PKI}'s performance with ``pre-quantum'' as well as quantum-safe cryptographic algorithms.
\end{abstract}


\section{Introduction}
\label{intro}
5G is the most recent standard for mobile communication and provides major enhancements compared to earlier cellular technologies. With 5G, high data transfer rates (up to 10 Gbps), very low latency (below one millisecond),  high-speed mobility (up to 500 km/h) and high connection densities (one million per square kilometer) are possible \cite{5g-netw-perf}.
The applications of 5G mainly belong to the following categories:
\begin{itemize}
	\item \textit{Ultra-Reliable and Low Latency Communications (URLLC)}- This use case appeals to industries such as healthcare, manufacturing that need fast and highly reliable data transmission.
	\item \textit{Enhanced Mobile Broadband (eMBB)}- Improving upon existing mobile broadband services, this use case focuses on higher data transmission speeds.
	\item \textit{Massive Machine-Type Communications (mMTC)}- This use case expands the scope of Internet-of-Things by adding even more devices to the network through supporting high density of devices in a particular region and low power consumption.
\end{itemize}

Since 4G/5G networks are IP-based, it opens them to a host of cyberattacks which were not possible with previous generation mobile networks. Though 3GPP standards have mandated several security features for 5G networks (Inter-operator security provided by security proxy servers, subscriber identifier encryption for privacy, mutual authentication for network and devices), implementation issues can introduce vulnerabilities in telecommunication networks. Sometimes the 3GPP standard specifications are themselves vulnerable to exploitation. Several such vulnerabilities have been discovered by academic and industry researchers. 

Cyberattacks against real-world 4G/5G telecommunication networks have already begun with Vodafone Portugal reporting an attack in February 2022 \cite{vodafone-attack} which affected its services such as the 4G/5G network, television and voice/digital answering services, for millions of customers. Additionally, ATM networks and public emergency services including ambulance operators, fire departments, and hospitals were also impacted. The telco company had stated that the recovery process would take time and involve multiple teams given the scale and seriousness of the attack.

The 3GPP 5G Service-based Architecture (SBA) security specifications Rel-15 and Rel-16 \cite{3gpp-sec-rel15, 3gpp-sec-rel16}, detail security for direct and indirect communication (through proxy) between Network Functions (NFs) in non-roaming and roaming scenarios. The communication is based on mutual authentication and transport security between NFs, using TLS 1.2 and 1.3. However, the 3GPP specifications leave several details on how to provision certificates and how to setup an appropriate Public Key Infrastructure (PKI) for 5G SBA, unspecified.

In this work, we present \textsf{5G-SBA-PKI}, a public key infrastructure for secure inter-NF communication in 5G cellular networks. 5G-SBA-PKI is designed to include multiple certificate authorities (with different scopes of operation and capabilities) at different PLMN levels for certification operations and key exchange between communicating NFs. We also incorporate distribution of trust over multiple entities (CAs and certificate transparency log servers), taking inspiration from existing web PKI. We conduct a formal analysis of 5G-SBA-PKI with respect to the desired security properties using TAMARIN prover \cite{tamarin-prover}. Finally, we evaluate \textsf{5G-SBA-PKI}'s performance in terms of processing times for certificate operations and inter-NF TLS handshake latency. The main contributions of our work are as follows: 
\begin{itemize}
	\item We design \textsf{5G-SBA-PKI}, a PKI for 5G SBA which supports secure inter-NF communication.
	\item We perform a formal security analysis of \textsf{5G-SBA-PKI}.
	\item We evaluate \textsf{5G-SBA-PKI}'s performance with ``pre-quantum'' as well as quantum-safe cryptographic algorithms.
\end{itemize}
 

\section{Related Work}
\label{literature}
There have been several proposals to design web PKI in literature. 
AKI \cite{aki} presents a certification infrastructure that handles adversarial events, such as the compromise of a CA’s private key, distributes trust by implementing a system of checks and balances among independent parties (Certification Authority, Integrity Log Server and validators) while efficiently managing common certificate operations. ARPKI \cite{arpki} is an attack-resilient PKI which is co-designed with a formal model. It offers substantially stronger security guarantees, formal machine-checked verification of its core security property and low overhead/additional latency to TLS (through proof-of-concept implementation). ARPKI uses three entities for the certificate operation: two CAs and an ILS, where the CAs conduct active on-line confirmations with validator-like capabilities. Yu et al. have proposed DTKI \cite{dtki}, a formally analysed PKI for web certificate management which reduces reliance on a fixed set of CAs, aims to secure web data transfer even if all the involved PKI entities are compromised, and formalises the underlying data structure used to store certificate logs. DTKI includes multiple certificate log maintainers, a mapping log maintainer, log mirrors and CAs. More recently, F-PKI \cite{fpki} allows varying levels of trust in CAs instead of trusting all CAs equally which is the case for today's PKI. Domain owners are allowed to set policies specifying the list of CAs authorized to issue certificates for their domains. Clients connecting to those domains can set a validation policy to express different levels of trust in the authorized CAs. 

A few works have conducted formal analysis of 5G authentication protocols which deal with the mutual authentication between mobile subscribers and 5G carrier networks. In \cite{5g-formal-analysis1}, authors have presented formal models of 5G AKA (Authentication and Key Agreement) and EAP-AKA protocols and used \textit{Tamarin} (a security protocol verification tool) to carry out an automated, systematic security evaluation of the models with respect to the 5G security goals. Similarly, \cite{5g-formal-analysis2} presents a formal model of the 5G EAP-TLS authentication protocol built using applied pi calculus and an automated security analysis of the model by \textit{Proverif} model checker. 

However, designing PKI for 5G cellular networks has not received much attention from the academic cybersecurity community. The closest work that we could find is \cite{distant}, in which the authors have presented DISTANT, a decentralized key management scheme for 5G mobile small cell networks. It combines threshold secret sharing with self-generated certificates where network nodes are provided proxy keys, allowing them to issue and sign certificates for themselves.

\textbf{Difference from State-of-the-Art PKIs}: The PKIs proposed in \cite{aki,arpki,dtki,fpki} are designed for web-based client-server TLS connections, not inter-NF communication in 5G SBA which is the focus of our work. Directly applying the web PKI proposals to 5G SBA would not address specific requirements of inter-NF communication in a 5G SBA, which are listed as follows: 
\begin{itemize}
	\item Inter-NF communication would require automated certificate management.
	\item The communicating NFs may be directly or indirectly connected (through proxy) and may be located in the same or different PLMNs (roaming).
\end{itemize}

DISTANT \cite{distant} addresses certificate operations, however, it is targeted at securing the communication between mobile devices and mobile small cell base stations with no aspect of inter-NF communication in 5G SBA taken into consideration.

\section{Background}
\subsection{5G Service Based Architecture}
The mobile cellular network enables wireless connectivity for devices such as smartphones and tablets, referred to as User Equipment (UE).
It is made up of two primary subsystems: the Radio Access Network (RAN) and the Mobile Core. The RAN is responsible for managing and ensuring efficient usage of radio resources such as spectrum and meeting users' Quality of Service (QoS) requirements. It consists of several base stations (each covering a particular area), known as eNB (evolved Node B) in 4G and gNB (next 'g'eneration Node B) in 5G. 
The Mobile Core is a collection of functions that provides several services including authenticating devices before allowing them to connect to the network, ensuring that the connectivity meets QoS requirements, providing IP connectivity for data-based services, tracking device mobility and tracking subscriber usage.

\begin{figure*}[t]
	\caption{5G Mobile Core Architecture (Source: Open Cloud \cite{open-cloud-github})}
	\label{5gc-arch}
	\centering
	\includegraphics[scale=0.5]{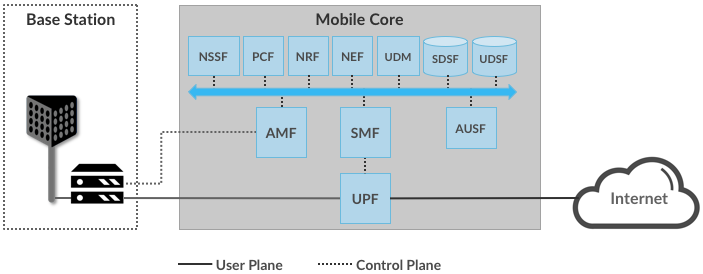}
\end{figure*}

The 5G Mobile Core (5GC) adopts a microservice-like architecture (shown in Fig. \ref{5gc-arch}), with a set of functional blocks, each of which is called a Network Function (NF). The NFs are divided into three groups, with two groups running in the Control Plane (CP) and one group running in the User Plane (UP). The block which provides the means for NFs to discover each other and related services is called the Network Repository Function (NRF).

\subsection{3GPP Security Specifications for 5G SBA}
3GPP TS 33.501 v16.3.0, Section 13 specifies security for indirect communication between NFs. When an NF consumer wants to use an NF producer's service, it does not communicate with it directly. Instead, it goes through a Service Communication Proxy (SCP) that acts as an intermediary between them. This affects the security aspect of the communication. The consumer and producer have to trust the SCP to relay their messages correctly. The SCP can also change the messages according to some standard or custom features. Therefore, even though each hop uses TLS for mutual authentication and transport security, the consumer and producer cannot have end-to-end transport security with TLS like they would in direct communication.

A new NF called security edge protection proxy (SEPP) was introduced in the 5G architecture by 3GPP SA3 (as shown in Fig. \ref{5g-pki-arch}). All signaling traffic across PLMNs is expected to transit through these security proxies. 
The interface (N32-f) enables secure communication between NF service consumers and producers located in different PLMNs. This security is provided by the SEPPs in both networks, hereafter referred to as \textit{cSEPP} and \textit{pSEPP} respectively. cSEPP and pSEPP are connected by interconnect providers they have business relationships with, called \textit{cIPX} and \textit{pIPX} respectively. Here, IPX refers to IP exchange service. cIPX and pIPX may have more interconnect providers between them, but they are expected to just forward the IP traffic.
 

\section{5G-SBA-PKI Design}
\label{system-overview}

\subsection{PKI Architecture}
\label{pki-arch}
Our proposed PKI, \textsf{5G-SBA-PKI}'s architecture is shown in Fig. \ref{5g-pki-arch} using two PLMNs, \textit{PLMN-1} and \textit{PLMN-2} for illustration purposes. The NFs in PLMN-1 and the NFs in PLMN-2 can communicate with each other through their respective SEPPs. 5G-SBA-PKI consists of the following entities:  

\begin{figure*}[t]
	\caption{\textsf{5G-SBA-PKI} Architecture}
	\label{5g-pki-arch}
	\centering
	\includegraphics[scale=0.5]{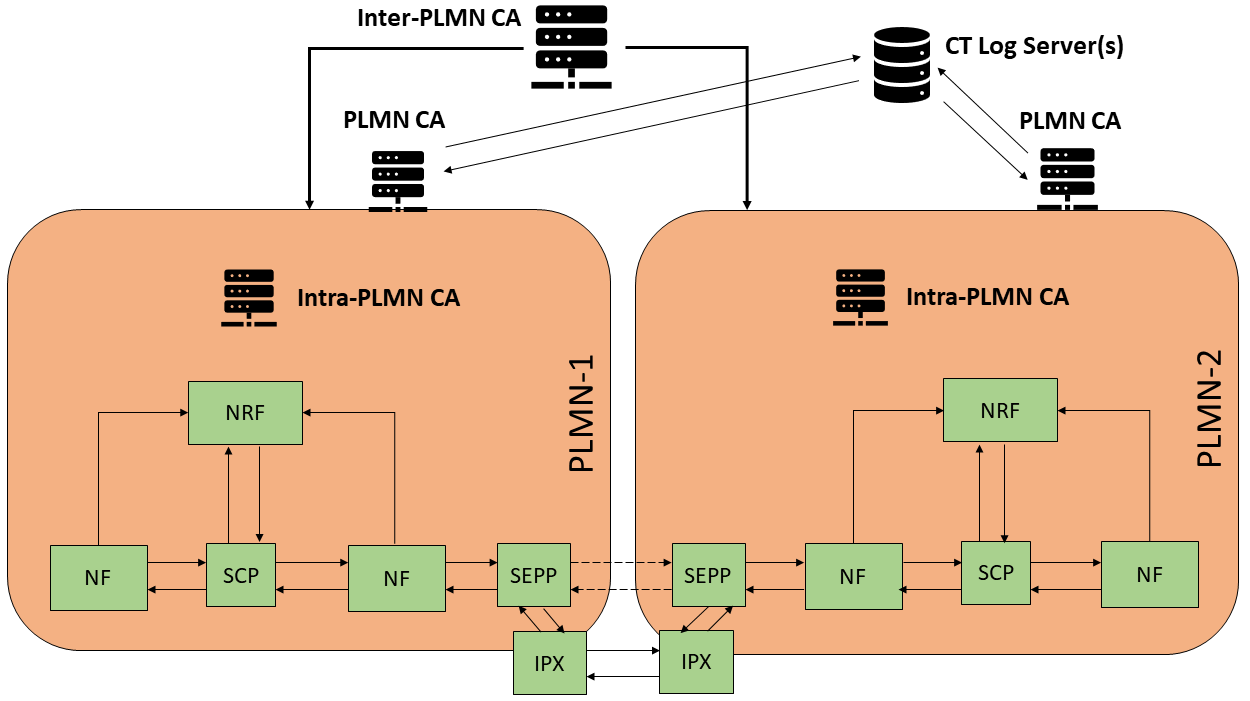}
\end{figure*}

\textbf{Intra-PLMN CA}: It issues and revokes certificates to NRF, NFs and SCPs with a certain expiry period. There is no certificate chain, i.e., intra-PLMN CA acts as root CA for all entities located inside a PLMN. Its operations are listed below:

\begin{enumerate}
	\item Automatic certificate generation and issue to NFs, SCPs, NRF
	\item Sharing of list of supported ciphers (symmetric encryption algorithms) with the NFs, SCPs, NRF
	\item Automatic certificate update, renewal (after checking that entity is behaving as expected)
	\item Certificate revocation
	\item Certificate revocation list (CRL) maintenance
\end{enumerate}

\textbf{PLMN CA}: It issues and revokes certificates for the IPX providers which the underlying PLMN has a business relationship with, the IPX providers' roaming partners and the SEPP which is part of the PLMN's core network. This is in line with the recommendations in 3GPP TS 33.501 v16.3.0, Section 13.2.4.5.2. SEPP supports TLS wildcard certificate for its domain name (3GPP TS 33.501 v16.3.0, Section 13.1.1.0). SEPPs exchange IPX certificates needed to verify IPX modifications at receiving SEPP (3GPP TS 33.501 v16.3.0, Section 13.2.2.2). The certificate chain for certificates issued to PLMN SEPPs is as follows: PLMN CA $\rightarrow$ Inter-PLMN CA $\rightarrow$ Root CA. PLMN CA's operations are listed below:

\begin{enumerate}
	\item Certificate generation and issue to IPX providers, roaming partners, SEPP on request
	\item Certificate update, renewal on request (after checking that entity is behaving as expected) 
	\item Certificate revocation
	\item CRL maintenance
\end{enumerate}

\textbf{Inter-PLMN CA}: It issues and revokes certificates to PLMN CAs. The certificate chain for certificates issued to PLMN CAs is as follows: Inter-PLMN CA $\rightarrow$ Root CA. The inter-PLMN CA's operations are listed below:

\begin{enumerate}
	\item Certificate generation and issue to PLMN CAs on request
	\item Certificate update, renewal on request
	\item Certificate revocation
	\item CRL maintenance
\end{enumerate}

\textbf{Certificate Transparency Log Server}: It maintains Certificate Transparency (CT) logs as standardised by IETF RFC 9162 \cite{cert-trans-rfc} for monitoring and auditing the issuance of SEPP certificates. CT logs are append-only logs such that the PLMN CA registers certificates issued to SEPPs with the CT log server. The CT logs are stored in a Merkel hash tree where the leaves contain certificates and root/intermediate nodes contain the hash of concatenated child nodes. An entity which wishes to verify the legitimacy of a certificate can query the CT log server which returns a non-repudiable audit proof for the certificate.

\subsection{PKI Operation}

When an NF/SCP/NRF in a PLMN core network comes online, the intra-PLMN CA automatically signs and issues a certificate to them after verifying their identity and that they are operating normally. The NF uses 5G-SBA-PKI certificate (\textit{5GCert}) to authenticate itself while connecting to another NF, the NRF, or SCP within the core network. The NRF uses 5GCert to authenticate itself while connecting to an NF or SCP within the core network. The SCP uses 5GCert to authenticate itself while connecting to an NF or the NRF within the core network. The validity period of the 5GCerts issued by intra-PLMN CA may be decided by individual PLMN operators. Once the validity period of a 5GCert issued to an entity is over, the intra-PLMN CA can automatically re-issue a 5GCert to that entity after verifying their identity and normal operation again. However, if any of the entities (NF/SCP) is detected as compromised (for example by an IDS installed in core network) or ceases to operate, the intra-PLMN CA can be immediately notified through the NRF. If the NRF itself is compromised, the intra-PLMN CA can be notified through the Policy Control Function (PCF). Once the intra-PLMN CA is notified, it may revoke the 5GCert immediately. The intra-PLMN CA also maintains a list of all entities whose 5GCerts have been revoked in the past.

When a PLMN operator forms a business relationship with an IPX provider to connect with other PLMNs, the PLMN CA signs and issues 5GCerts to the IPX provider as well as the provider's roaming partners after verifying their identity. The IPX provider having a relationship with a PLMN uses the 5GCert to authenticate itself to the IPX provider having a relationship with another PLMN when NFs in the two PLMNs need to communicate. The validity period of the 5GCerts issued by PLMN CA may be decided by individual PLMN operators. If an IPX provider's details change or is compromised, the PLMN CA can update or revoke the 5GCert respectively once it is notified.

When an SEPP in a PLMN comes online, it registers its domain with the PLMN CA and requests the CA to sign its certificate. The PLMN CA is expected to have already obtained a certificate signing its public key from the inter-PLMN CA. The PLMN CA first verifies the identity of the SEPP and then forms a chain by appending its own certificate with the public key of the requesting SEPP and signing the resulting wildcard 5GCert (in line with 3GPP TS 33.501 v16.3.0, Section 13.1.1.0) with a certain validity period which it forwards to the SEPP. For example, if $K_{SEPP}$ is the public key of requesting SEPP, $K_{PLMN-CA}$ is the public key of the PLMN CA, then the certificate issued to PLMN CA by inter-PLMN CA, \textit{PLMN-CA-Cert} would be: $\{PLMN-CA, K_{PLMN-CA}\}_{K_{Inter-PLMN-CA}^{-1}}$ and the 5GCert would be: $\{SEPP, K_{SEPP}, PLMN-CA-Cert\}_{K_{PLMN-CA}^{-1}}$. The PLMN CA registers the SEPP's 5GCert with one or multiple CT log servers. Each CT log server adds the SEPP domain's 5GCert to its database and re-computes the hash values over all stored certificates. The SEPP receives the audit proof downloaded from every CT log server along with its 5GCert from PLMN CA and uses it to authenticate itself while connecting to another SEPP or when another SEPP is requesting a connection. The receiving SEPP uses the trusted root-CA certificates to validate the 5GCert and the pre-installed CT log server public keys to validate the audit proof. The concise message flows during above 5G-SBA-PKI operations have been shown in Fig. \ref{5gpki-msg-flows}. 

We recommend the inter-PLMN CAs to be different than the Internet CAs for security and efficiency reasons. Since the PLMN networks offer much more critical services than web domains, having common CAs for both PLMNs and web domains means that if the CA is compromised, both the web domains and PLMN domains registered with that CA can be subject to cyber attacks. Further, if the same CA services both web domains as well as PLMN domains, the CA's responses might be slowed down especially during periods of busy network traffic. 



\begin{figure}[h]
	\caption{Message Flows During 5G-SBA-PKI Operations}
	\label{5gpki-msg-flows}
	\footnotesize
	\setlength{\tabcolsep}{10pt}
	\noindent\rule{\linewidth}{0.5pt}

	\begin{description}
		\item[\textbf{5GCert Issuance to NFs/SCP/NRF}]
	\end{description}
	
	\textit{Intra-PLMN CA} $\rightarrow$ \textit{X} : Automatically send signed 5GCert, $\{X, K_{X}\}_{K_{Intra-PLMN-CA}^{-1}}$ 	
	
	\begin{description}
		\item[\textbf{5GCert Issuance to SEPP}]
	\end{description}
	
	\textit{SEPP} $\rightarrow$ \textit{PLMN CA} : Register domain, request to sign certificate \\
	\textit{PLMN CA} : Verify SEPP's identity \\
	\textit{PLMN CA} $\rightarrow$ \textit{SEPP} : Send signed 5GCert, $\{SEPP, K_{SEPP}, PLMN-CA-Cert\}_{K_{PLMN-CA}^{-1}}$; CT audit proof
	
	\begin{description}
		\item[\textbf{SEPP 5GCert Registration}]
	\end{description}
	
	\textit{PLMN CA} $\rightarrow$ \textit{CT log server(s)} : Register SEPP domain's 5GCert \\
	\textit{CT log server} : Add SEPP domain's 5GCert, re-compute hash values \\
	\textit{CT log server(s)} $\rightarrow$ \textit{PLMN CA} : Send audit proof
	
		\begin{description}
		\item[\textbf{TLS Connection (Within PLMN)}]
	\end{description}
	
	\textit{NF-C} $\rightarrow$ \textit{NF-P} : TLS connection request, $5GCert_{NF-C}$ \\
	\textit{NF-P} $\rightarrow$ \textit{NF-C} : $5GCert_{NF-P}$ \\

	\begin{description}
		\item[\textbf{TLS Connection (Between PLMNs)}]
	\end{description}
	
	\textit{cSEPP} $\rightarrow$ \textit{pSEPP} : TLS connection request, $5GCert_{cSEPP}$, CT audit proof \\
	\textit{pSEPP} $\rightarrow$ \textit{cSEPP} : $5GCert_{pSEPP}$, CT audit proof
	
	\noindent\rule{\linewidth}{0.5pt}
\end{figure}

\section{Security Analysis of 5G-SBA-PKI}

\subsection{TAMARIN Background and Modelling}
We formally analyze 5G-SBA-PKI's security properties using TAMARIN prover \cite{tamarin-prover}, which is a cutting-edge tool for analyzing and verifying protocols. Protocols are represented using multiset rewriting rules, while properties are expressed using a subset of first-order logic that allows for quantification over timepoints. TAMARIN can automatically verify many cases, but also allows for manual proof tree traversal for interactive verification. If no proof is found, the tool provides a counter-example. More details about the TAMARIN prover and how it works can be found at \cite{tamarin-manual}. We used the communication flow of 5G-SBA-PKI as illustrated in Fig. \ref{5gpki-msg-flows} as a reference while modelling.

\subsection{Adversary Model}
TAMARIN adheres to the Dolev-Yao model, in which the adversary has the ability to intercept and block all messages, as well as view and alter messages (or parts of them) that are not secured through cryptography. In TAMARIN, the operation of a security protocol in the presence of an adversary is represented using labelled multiset rewriting rules. Additionally we assume that the adversary can compromise any of the CAs and CT log server.

\subsection{Proof Goals}
\label{proof-goals}
\textit{Goal \#1}: When the communicating NF-C/NF-P have been issued their respective 5GCerts, and the NF-C/NF-P reciprocally accept connections to each other directly or through an SCP, the adversary does not know the private keys for NF-C, SCP or NF-P.

\textit{Goal \#2}: When the communicating cSEPP/pSEPP have been issued their respective 5GCerts, and the cSEPP/pSEPP reciprocally accept connections to each other, the adversary does not know the private keys for cSEPP or pSEPP.  

\begin{figure*}[t]
	\caption{TAMARIN Model Security Property \#2}
	\label{tamarin-lemma-1}
	\centering
	\includegraphics[scale=0.5]{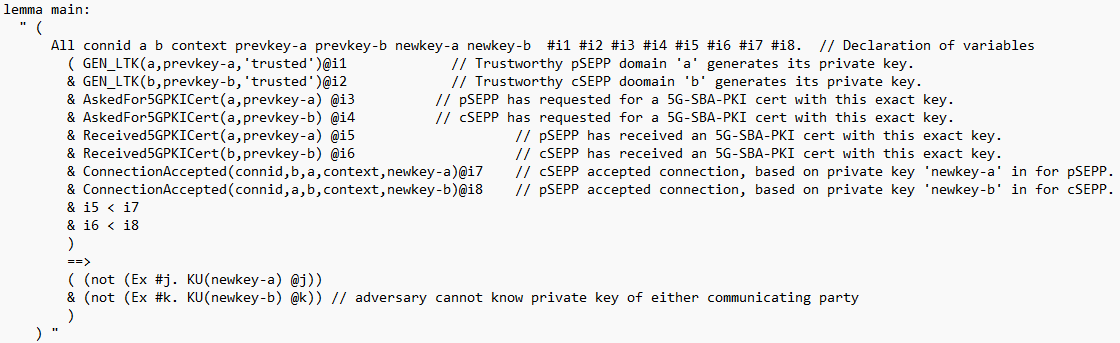}
\end{figure*}

We analyze goal \#1 for two cases: when none or all the involved entities (i.e., intra-PLMN CA) are compromised while goal \#2 is analyzed for two cases: with at most two entities compromised, and with all entities (PLMN CA, inter-PLMN CA, CT log server) compromised. As an example, the lemma that encodes proof goal \#2 in TAMARIN is shown in Figure \ref{tamarin-lemma-1}. The formula starts with a quantification over variables ($connid$, $a$, $b$, $\dots$): for all values of those variables there should be a GEN\_LTK($\dots$) action in the trace at position $i1$ ($\#i1$ denotes a variable $i1$ of type `timepoint'). In our model, this action can only be produced by a particular rule that generates initially trusted keys.

\subsection{Analysis Results}
The security goal \#1 is only achieved when none of the applicable entities (intra-PLMN CA) are compromised otherwise TAMARIN produces a counter-example attack since an intra-PLMN CA controlled by adversary can create 5GCert for any NF within a PLMN which is intuitive. The security goal \#2 is achieved for at most two compromised entities out of (PLMN CA, inter-PLMN CA, CT log server), i.e., TAMARIN verifies the lemma. For all three entities compromised, TAMARIN finds the expected attack since adversary can create 5GCert for any SEPP domain. We ran our experiments on a VMWare ESXi server VM with Intel Xeon Silver 4216 CPU @2.10GHz, 64-bit architecture, 4 cores, 16GB RAM and running Ubuntu 20.04. The CPU timings taken for TAMARIN to verify security goals \#1 and \#2 are shown in Table \ref{tamarin-cpu-time}.

\begin{table}[]
	\centering
	\begin{tabular}{|l|l|l|l|l|}
	\hline
	\textbf{Security Goal} & \multicolumn{3}{|l|}{\textbf{CPU Time (in milliseconds)}} \\ \cline{2-4}
	 & \thead{No\\ compromised\\ entity} & \thead{Max. two\\ compromised\\ entities} & \thead{All\\ entities\\ compromised} \\ \hline
	Goal \#1 &  340 & - & 490\\ \hline
	Goal \#2 & - & 656 & 627\\ \hline
	\end{tabular}
	\caption{CPU timings for TAMARIN's verification of security lemmas}
	\label{tamarin-cpu-time}
\end{table}

\subsection{Formal Analysis Comparison with State-of-the-Art}
ARPKI\cite{arpki} and DTKI \cite{dtki} are the only state-of-the-art web-based PKI schemes which have been formally analyzed using automated tools such as TAMARIN prover. The main security property verified during ARPKI's formal analysis is that when a web domain has requested and obtained an ARPKI certificate (ARCert), and a browser accepts a connection to said domain with the ARCert, then an adversary with full-network control can not know the ARCert's private key when at most two entities have been compromised. The security properties proven for DTKI are: (1) when all service providers (i.e. CAs, the MLM and CLMs) are compromised, and the domain owner has successfully verified his master certificate in the log, then the attacker cannot learn the message exchanged between a user and a domain owner, and (2) when all service providers
are compromised, and a domain owner has not verified their master certificate, and the attacker learns the message exchanged between a user and the domain owner, then afterwards the domain owner can detect this attack by checking the log. The security properties verified during 5G-SBA-PKI's formal analysis as mentioned in goals \#1 and \#2 in sub-section \ref{proof-goals} are similar to above where the NF-P/pSEPP is analogous to the web domain while the NF-C/cSEPP are analogous to the browser/user. Similar to ARPKI, we prove the desired security guarantees for at most two and all compromised entities.

However, there are differences that arise during the formal modelling of 5G-SBA-PKI due to inherent differences in the architecture, components and application scenario as compared to ARPKI and DTKI. Unlike the web domain, NFs do not request for a certificate from a CA, rather they are issued certificates (5GCerts) automatically by the intra-PLMN CA. Further, the entities involved in ARPKI (two CAs, one ILS) and DTKI (CAs, the MLM and CLMs) are different from the entities involved in 5G-SBA-PKI (intra-PLMN CA, PLMN CA, inter-PLMN CA, CT log server). In case of ARPKI/DTKI, the web browser authenticates the domain based on its ARCert/TLS certificate and accepts the connection while in case of 5G-SBA-PKI, both the NF-C/cSEPP and NF-P/pSEPP authenticate each other mutually based on their respective 5GCerts and accept the connection towards each other.

\subsection{AutomatedDropWithTimer Attack}
Coldwell et al. \cite{5gad} have provided a dataset, 5GAD-2022 which includes ``normal'' network traffic packets (various automated tasks such as downloading files, streaming online videos, accessing websites were run on single and multiple UEs) and ``attack'' packets (attacks such as reconnaissance, network reconfiguration, DoS were targeted at a 5G core network). The packets were captured on four different network interfaces within the 5G core (N2, N3, N4, N5). An interesting DoS attack, \textit{automatedDropWithTimer}, has been represented in the dataset which works as follows. The attacker listens to network traffic and waits for a request to establish a packet forwarding control protocol (PFCP) session while the UE is connecting to a 5G network. On receiving the request, it checks if the IP address of UE belongs to a list containing IP addresses of targeted victims and if found, the PFCP session information is recorded and subsequently used for UE traffic redirection. The attacker machine/VM posing as a valid SMF, sends a session modification request with the session ID extracted from the recorded session information and the forwarding action rule ID (FARID) to the user plane function (UPF). By alternating between user traffic redirection and dropping, the attacker forces the user to be disconnected from the data network, thus denying it Internet access. 

Here, PFCP is a protocol used on the N4 interface between the control plane and the user plane, and it helps the SMF establish a session on the UPF to manage the GTP tunnel that provides Internet access to the subscriber. Mutual authentication based on TLS between the SMF and UPF using 5GCert can help prevent such attacks. An attacker SMF would not be able to present a valid 5GCert to the UPF during the mutual authentication phase of the TLS connection setup, leading to denial of connection request and hence, the attacker SMF would not be able to connect to the UPF in the first place. 

\section{Implementation}
We built a 5G test environment using Open Air Interface (OAI) software \cite{oai-softw} which includes Docker containers corresponding to the various 5G core network components (e.g., AMF, SMF, NRF) as well as the OAI gNB and NR-UE emulators. We built separate container images to implement the SCP and SEPP since they have not been implemented in the current release of OAI 5G core network. The Docker container images were deployed on a VMWare ESXi server VM with Intel Xeon Silver 4216 CPU @2.10GHz, 64-bit architecture, 8 cores, 16GB RAM and running Ubuntu 18.04. We also tested Internet connectivity from the NR-UE successfully. We used the \textit{OpenSSL} library \cite{openssl-github} to implement the CA components (intra-PLMN CA, PLMN CA, inter-PLMN CA) and Google Certificate Transparency Go \cite{cert-trans-github} to implement the CT log server component of our proposed \textsf{5G-SBA-PKI} in the 5G test environment.

\subsection{Performance Evaluation}
In Table \ref{avg-proc-time}, we have shown the average processing time (in seconds) for two certificate operations- certificate request (CertReq) and signing (CertSign). The time taken for each certificate operation has been averaged over 1000 test runs. The average processing time for each certificate operation has been shown for multiple encryption algorithms- RSA 2048-bit, Elliptic curve-cryptography (ECC): P-256, and Quantum-resistant (QR) algorithms: Dilithium-2. The \textit{OpenSSL} library was used to test RSA and ECC algorithms while the Open Quantum Safe (OQS) \cite{open-quantun-safe-github} library was used to test quantum-safe algorithms. The average processing time for CertReq for RSA-2048 is two orders of magnitude greater than that for ECC and QR algorithms. Further, CertReq takes longer time than CertSign for RSA-2048 (almost 25 times) but the opposite is true for QR algorithms.

\begin{table}[]
	\centering
	\begin{tabular}{|l|l|l|l|l|}
	\hline
	\textbf{CA Operation} & \multicolumn{3}{|l|}{\textbf{Avg. Processing Time (in secs)}} \\ \cline{2-4}
	 & \textbf{RSA} & \textbf{P-256} & \textbf{Dilithium2} \\ \hline
	CertReq &  0.10469 & 0.003073 & 0.00312\\ \hline
	CertSign & 0.00439 & 0.00296 & 0.00362\\ \hline
	\end{tabular}
	\caption{Average processing time for certificate operations}
	\label{avg-proc-time}
\end{table}

\begin{figure}[h]
	\caption{Cumulative Distribution Function of CertReq processing times}
	\label{cdf-certreq-proc-time}
	\centering
	\includegraphics[scale=0.5]{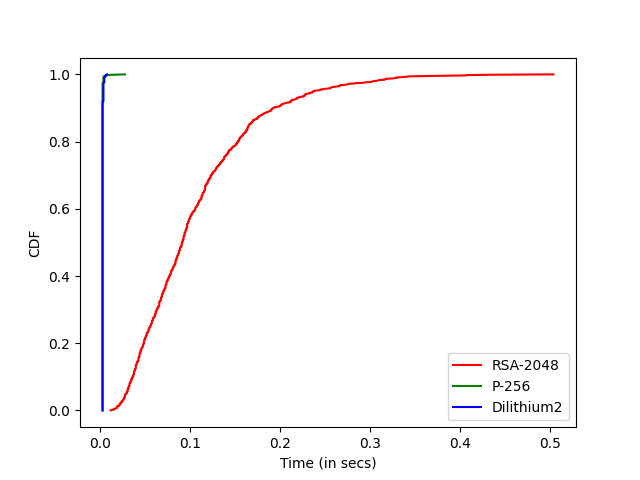}
\end{figure}

\begin{figure}[h]
	\caption{Cumulative Distribution Function of CertSign processing times}
	\label{cdf-certsign-proc-time}
	\centering
	\includegraphics[scale=0.5]{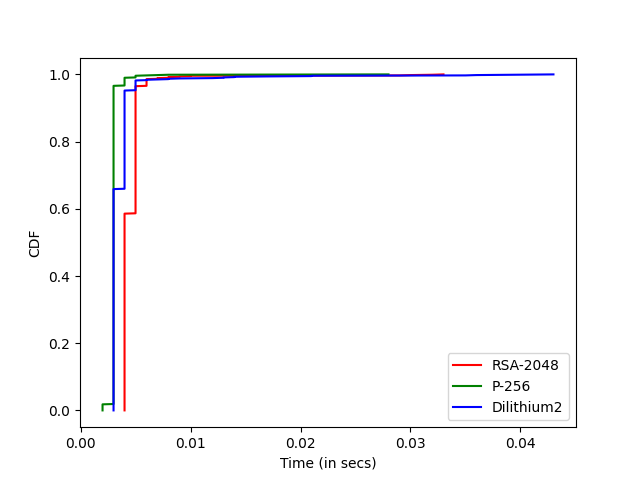}
\end{figure}

We have also plotted the CDF (Cumulative Distribution Function) of the CertReq and CertSign processing times in Fig. \ref{cdf-certreq-proc-time} and Fig. \ref{cdf-certsign-proc-time} respectively for RSA-2048, ECC and QR algorithms mentioned above using 1000 samples for each distribution. The CDF for CertReq operation using RSA-2048 resembles a $\chi^2$ distribution whereas the CDFs for CertSign operation resembles a Bernoulli distribution. 

Next, we set up a TLS server running on the AMF using the certificates and keys generated through \textit{openssl} utility with RSA-2048. A TLS client running on SMF connected to the TLS server to emulate inter-NF communication within a PLMN. We calculated the average RAM utilized at AMF and SMF for the TLS connection (over 100 test runs) by calculating the difference between the RAM utilized before and after the TLS connection was set up. The results are shown in Table \ref{avg-ram-tls-conn}. It can be seen that on an average, TLS connection takes up more than 4 times RAM at the NF consumer than at the NF producer. However, given that the RAM in production environment containers is usually in gigabytes, the percentage of RAM utilized during TLS connection set up is quite small and should not affect the main NF operations. We also calculated the average time taken for the TLS handshake between AMF and SMF (over 100 test runs) and it was found to be \textit{2 milliseconds} which is quite low compared to the average TLS handshake time period for web client-server connections (250ms - 500ms) \cite{tls-handshake-latency}. Since 5G targets sub-millisecond latency, the TLS handshake as well as the 5G-SBA-PKI certificate operations may need to be optimized further.

\begin{table}[]
	\centering
	\begin{tabular}{|l|l|}
	\hline
	\textbf{Object} & \textbf{Avg. RAM utilized} \\ \hline
	{NF producer} & 995kb \\ \hline
	{NF consumer} & 4.26Mb \\ \hline
	\end{tabular}
	\caption{Average RAM utilized for NFs during TLS connection}
	\label{avg-ram-tls-conn}
\end{table}

\section{Discussion}
\subsection{Performance Comparison with State-of-the-Art}
Most state-of-the-art PKI proposals such ARPKI, TriPKI use OpenSSL library for digital signatures as does 5G-SBA-PKI. Hence, there is no need for a comparison of processing times for certificate operations among them. 
The performance of ARPKI, TriPKI and similar existing PKI proposals has been evaluated in terms of the average processing times for operations such as registration request, update request, confirmation request. If we have to perform any meaningful comparison between the performance of ARPKI and our proposed 5G-SBA-PKI, we would have to map ARPKI operations to those in 5G-SBA-PKI. Taking the example of ARPKI in specific, registration request is a message sent by a web domain owner to $CA_1$ and then by $CA_1$ to ILS to designate the trusted entities ($CA_1$, $CA_2$, ILS) during the initial registration of \textit{ARCert} certificate issued by ARPKI, where CA stands for Certificate Authority and ILS stands for Integrity Log Server. Update request messages are similar to the registration request messages and are sent when domain owner is updating the ARCert. Confirmation request is a message sent by domain to the ILS through $CA_1$ after at most every $ILS_{TIMEOUT}$ period, and in response a fresh confirmation (that the domain’s ARCert is valid) is sent by ILS to the domain which it then forwards to the client application. The focus of ARPKI as well as its architecture are very different from those of our proposed 5G-SBA-PKI. ARPKI focuses on improving AKI \cite{aki} for secure web client-server communication by introducing CAs with validator like capabilities so that there are no separate validators. 5G-SBA-PKI instead focuses on securing inter-NF communications in a 5G PLMN as well as across PLMNs by introducing CAs at various PLMN levels, it has no validator CAs or ILS as part of its architecture. Thus, it is not possible to adapt ARPKI operations to those in our proposed 5G-SBA-PKI. 

\subsection{Support for Tenant PKIs and Multi-Vendor NFs}
NFs in 5G SBA are typically realised as containerized applications with the containers orchestrated using one or more tenants, e.g., Docker or Kubernetes. 5G network operators have two options for issuing certificates used in mutual authentication of NF containers:
\begin{itemize}
	\item Use certificates generated by Intra-PLMN CA
	\item Use certificates issued by tenant-generated root CA
\end{itemize}

However, all NF containers may not be orchestrated using the same tenant. Multi-vendor 5G core networks is a paradigm where multiple vendors are used to provide NFs for a 5G core. For example, \textit{Vendor 1} can provide NFs related to subscriber data management, \textit{Vendor 2} can provide NFs related to signalling/routing and network slicing, and so on. This approach has many advantages such as less dependency on a single supplier, more network resiliency, ability to select the best supplier for supporting each service type and rapid deployment of services. The vendors which provide NFs may use different tenants to orchestrate the NF containers where each tenant may have their own PKI.

If the 5G network operator uses certificates issued by tenant-generated root CAs, and two NF containers belonging to different tenants have to authenticate each other, the two NFs would not be able to verify each other's certificates since those certificates would not have been signed by a common CA whom both the NFs could trust. Therefore, the network operator should only use certificates generated by a common Intra-PLMN CA.


\section{Conclusion}
We have proposed a public key infrastructure for 5G SBA core network, \textsf{5G-SBA-PKI} which ensures secure communication between network functions and between network functions and other entities. It consists of three CAs (at the intra-PLMN, PLMN and inter-PLMN levels) and a CT log server. We described the operation of \textsf{5G-SBA-PKI} entities and conducted its formal analysis based on desired security properties using TAMARIN prover. A performance evaluation of \textsf{5G-SBA-PKI} with ``pre-quantum'' as well as quantum-safe cryptographic algorithms reveals low average processing times for certificate operations on NF containers, low RAM utilization and low average handshake duration during TLS connection set up between NFs. 


\bibliographystyle{ieeetran}
\begingroup
\raggedright
\bibliography{5gpki}
\endgroup

%

\end{document}